\def\breakon{\end{multicols}\widetext\vspace{-.2cm}
\noindent\rule{.48\linewidth}{.3mm}\rule{.3mm}{.3cm}\vspace{.0cm}}
\def\breakoff{\vspace{-.2cm}
\noindent
\rule{.52\linewidth}{.0mm}\rule[-.27cm]{.3mm}{.3cm}\rule{.48\linewidth}{.3mm}
\vspace{-.3cm}
\begin{multicols}{2}
\narrowtext}
\documentstyle[aps,prl,multicol,epsfig]{revtex}

\begin{document}

\draft

\widetext

\title{Non-perturbative saddle point for the effective action of
disordered and interacting electrons in 2D}

\author{Claudio Chamon$^a$ and Eduardo R. Mucciolo$^b$} 

\address{ $^a$ Department of Physics, Boston University, Boston, MA
02215 \\ $^b$ Departamento de F\a'{\i}sica, Pontif\a'{\i}cia Universidade
Cat\'olica do Rio de Janeiro, \\
Caixa Postal 38071, 22452-970 Rio de Janeiro, 
Brazil}

\date{June 19, 2000}

\maketitle

 
\begin{abstract}
  We find a non-perturbative saddle-point solution for the non-linear
  sigma model proposed by Finkelstein for interacting and disordered
  electronic systems. Spin rotation symmetry, present in the original
  saddle point solution, is spontaneously broken at one-loop, as in
  the Coleman-Weinberg mechanism. The new solution is singular in both
  the disorder and triplet interaction strengths, and it also
  explicitly demonstrates that a non-trivial ferromagnetic state
  appears in a theory where the disorder average is carried out from
  the outset.
\end{abstract}




\begin{multicols}{2}

\narrowtext

Understanding the combined effects of disorder and interactions in
electronic systems has proven to be both an extremely interesting and
difficult problem. The issue has regained a new surge of interest
since the discovery by Kravchenko and co-workers \cite{Kravchenko} of
a possibly conducting state in two-dimensional (2-D) Si-MOSFETs (see
also Ref.~\onlinecite{others}). Although there has been intense debate
within the theoretical community as on the origin of the transition
\cite{everybody}, little has been accomplished that is in as solid
grounds as the scaling theory of localization for the non-interacting
problem \cite{Gang4}.

The initial attempt to establish a scaling theory for the interacting
problem was put forward by Finkelstein \cite{Finkelstein}, who studied
an extension of Wegner's non-linear sigma model containing singlet and
triplet interaction couplings. The renormalization group (RG) flow
takes the interaction coupling constants to strong coupling, away from
the perturbative starting point of a diffusive Fermi liquid state;
hence no conclusive picture has emerged from this approach. Among the
outstanding theoretical issues is the nature of the magnetic state
signaled by the divergence of the triplet interaction.

Let us begin by briefly discussing the breakdown of the RG flow for
the interaction. The one loop RG equation for $\gamma_2$, the ratio
between the triplet and the singlet interactions, is given by
\cite{Finkelstein,CCLM,Belitz-Kirkpatrick}
\begin{equation}
\frac{d{\gamma_2}}{dl} = \frac{1}{2}\, t\,
{\left(1+{\gamma_2}\right)^2}\ ,
\label{eq:fink-rg3}
\end{equation}
where $t$ is the resistance. Although the resistance also gets
renormalized, it is instructive to solve this equation for a ``fixed''
$t$, which leads to a breakdown at an RG scale $l = \frac{2}
{t(1+\gamma_2)}$, or at a length scale $\lambda = \Lambda^{-1}\,
e^{\frac{2} {t(1+\gamma_2)}}$ ($\Lambda$ is a momentum cutoff).

It is very useful to draw an analogy to the problem of BCS
superconductivity at this point. Within Shankar's renormalization
group approach to fermions \cite{Shankar}, one can perturb around the
Fermi liquid fixed point, and obtain the flow equation for the BCS
interaction $V<0$:
\begin{equation}
\frac{dV}{dl} = -V^2\ .
\end{equation}
This equation also breaks down at a finite length scale -- the
coherence length. At this length scale, the non-perturbative physics
of pairing takes over. Similarly, the breakdown of
Eq.~(\ref{eq:fink-rg3}) implies that non-perturbative physics
dominates the behavior of the system. Such physics cannot be accessed
perturbatively from the diffusive saddle point.

However, the main difference between these two examples is that, in
the case of BCS superconductivity, we know what the physics of the
non-perturbative fixed point is from the BCS mean-field
solution. Shankar's RG procedure allows us to find that there is an
instability of the Fermi liquid, but it alone does not access the
non-pertubative BCS solution. The situation is completely analogous in
the diffusive Fermi liquid problem: we know that the non-interacting
saddle point is unstable under the RG flow, but we cannot determine
the non-perturbative physics solely from the flow.

In this paper, we find a non-perturbative self-consistent solution for
the saddle point of the interacting and disordered electronic problem
by looking at the one-loop effective potential of the non-linear sigma
model. The new saddle point spontaneously breaks spin rotation symmetry
for {\it any} value of the triplet coupling constant. 

The starting point of our calculation is Finkelstein's $Q$-matrix model of
unitary class\cite{Finkelstein}. The disorder-averaged $N$-replica partition
function of the interacting problem reads
\begin{equation}
\langle Z_N \rangle = \int DQ \: e^{-S[Q]} ,
\end{equation}
where
\begin{eqnarray}
S[Q] & = & \frac{\pi \nu_F}{4\tau}\, {\rm tr}\, Q^2 - {\rm tr} \ln\,
G^{-1} + Q {\hat \gamma} Q ,
\end{eqnarray}
with the tensor ${\hat \gamma}$ containing the singlet and triplet
interactions, and
\begin{equation}
G^{-1} = i\omega + \left( \frac{\Delta}{2m} + \mu \right) +
\frac{i}{2\tau}\, Q.
\label{eq:Gfunction}
\end{equation}
The $Q$ matrix considered here has the following structure: $Q =
Q^{\alpha\beta}_{ij}(k - k^\prime; \omega,\omega^\prime)$, where the
pairs $ij$ and $\alpha\beta$ denote replica and spin indices,
respectively, while $k,k^\prime$ are momenta and
$\omega,\omega^\prime$ are frequencies. The trace operation runs over
all these indices. For our purposes, it will be sufficient to consider
only the zero-temperature limit.

Let us expand the action for $Q = Q_0 + \delta Q$:
\begin{eqnarray}
\label{eq:expansion}
S[Q] & = & S[Q_0] \nonumber \\ & + & \frac{\pi \nu_F}{2\tau}\, {\rm
tr}(Q_0\, \delta Q) - \frac{i}{2\tau}{\rm tr} (G_0\, \delta Q) + 2\,
Q_0 {\hat \gamma} \delta Q \nonumber \\ & + & \frac{\pi
\nu_F}{4\tau}\, {\rm tr}(\delta Q\, \delta Q) - \frac{1}{8\tau^2}\;
{\rm tr} (G_0\, \delta Q\, G_0\, \delta Q) - \delta Q {\hat \gamma}
\delta Q \nonumber \\ & + &\frac{i}{24\tau^3} \,
{\rm tr} (G_0\, \delta Q\, G_0\, \delta Q\, G_0\, \delta Q) ,
\end{eqnarray}
where $G_0^{-1}$ follows from Eq.~(\ref{eq:Gfunction}) after replacing
$Q$ by $Q_0$. The reason for keeping up to order $\delta Q^3$ when
searching for the saddle point will become transparent below. After
this expansion, the usual next step is to choose $Q_0$ such that the
term linear in $\delta Q$ vanishes; this leads to the saddle point
used by Finkelstein. Equivalently, this condition on the linear terms
can be cast as
\begin{equation}
\label{eq:vertex1-0} 
\frac{\delta S}{\delta Q}{\Big |}_{Q_0} = \Gamma^{(1)}_0{\Big
|}_{Q_0} = 0 ,
\end{equation}
where $\Gamma^{(1)}_0$ is the tree level (or zero-loop) one-vertex
function.  By writing the usual saddle-point equation as a condition
on the one-vertex $\Gamma^{(1)}_0$ (or linear in $\delta Q$)
potential, we are approximating the whole vertex function or effective
potential $\Gamma[Q]=V[Q]$ by the tree level potential $\Gamma_0[Q] =
S[Q]$.  This leads to the tree level saddle point
${Q_0}^{\alpha\beta}_{ij} (q; \omega, \omega^\prime) = {\rm sgn}
(\omega)\, \delta_{q,0}\, \delta_{\omega \omega^\prime}\,
\delta_{\alpha\beta}\, \delta_{ij}$. Fluctuations are then
parametrized by slow unitary rotations $Q = U^{-1}(r)\, Q_0\,
U(r)$. The sigma model constrained to this manifold has an instability
in the triplet interaction channel, which signals that the wrong
saddle point has been chosen.

\begin{figure} 
\hspace{.5 in}
\epsfxsize=1.6in
\vspace{.5cm}
\epsfbox{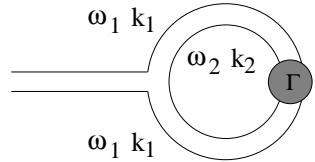} 
\caption{One-loop contribution to the one-vertex term in the 
  effective potential.}
\label{fig:loop} 
\end{figure}

We will proceed by considering not the tree level potential or action
in the derivation of saddle point, but the effective potential
generated by the fluctuations. There is a symmetry breaking not
present at the tree (or ``classical'') level, but that surfaces at
higher loop order (``quantum'') in the potential, an effect know as
the Coleman-Weinberg mechanism \cite{Coleman-Weinberg}. With this in
mind, the correct saddle for the interacting disorder electron problem
should be found at one-loop order:
\begin{equation}
\label{eq:vertex1-1} 
\frac{\delta V}{\delta Q}{\Big |}_{Q_0} = 
\Gamma^{(1)} {\Big |}_{Q_0} \approx \Gamma^{(1)}_0 {\Big |}_{Q_0}  +
\Gamma^{(1)}_1 {\Big |}_{Q_0} = 0 .
\end{equation}
This is the reason why we kept up to the cubic term in
Eq.~(\ref{eq:expansion}). To one-loop order, the cubic term
contributes to $\Gamma^{(1)}_1$, as shown in
Fig~\ref{fig:loop}. Notice that, if it were not for the interactions,
this contribution would vanish because it is proportional to $N\to 0$
in the zero-replica limit. The interaction keeps the contribution
alive because it requires all replicas in the loop to be the same.


Contracting two of the $\delta Q$'s from the cubic term, we obtain
(spin and replica indices are not shown):
\breakon
\begin{eqnarray}
\delta V & = & \frac{i}{8\tau^3} \sum_{\omega_1,\omega_2,\omega_3}\,
\sum_{k_1,k_2,k_3}\, G_0(\omega_1,k_1)\, G_0(\omega_2,k_2)\,
G_0(\omega_3,k_3)\, \langle \delta Q_{\omega_1 \omega_2}(k_1-k_2) \,
\delta Q_{\omega_2 \omega_3}(k_2-k_3) \rangle \, \delta Q_{\omega_3
\omega_1}(k_3-k_1)
\end{eqnarray}
or, alternatively,
\begin{eqnarray}
\Gamma^{(1)}_1 & = & \frac{i}{8\tau^3} \sum_{\omega_2}\,
\sum_{k_1,k_2}\, \left[ G_0(\omega_1,k_1) \right]^2 \,
G_0(\omega_2,k_2)\, \langle \delta Q_{\omega_1 \omega_2}(k_1-k_2) \,
\delta Q_{\omega_2 \omega_1}(k_2-k_1) \rangle.
\end{eqnarray}
\breakoff
\noindent
The propagator of density fluctuations ${\cal D}= \langle \delta Q
\,\delta Q \rangle $ follows from the quadratic part of
Eq.~(\ref{eq:expansion}) and depends on the saddle $Q_0$ through
$G_0$. Redefining the frequencies and momenta in the sums, one has
\begin{equation}
\Gamma^{(1)}_1 = \frac{i}{8\tau^3} \sum_{\omega^\prime} \, \sum_{k,q}
\, \left[ G_0(\omega,k) \right]^2 G_0(\omega + \omega^\prime,k + q)
\, {\cal D}(\omega,q)
\end{equation}
Next, we use the above $\Gamma^{(1)}_1$ and find the new saddle
solution of Eq.~(\ref{eq:vertex1-1}). The solution is a matrix $Q_0$
homogeneous in space and diagonal in the replica, spin, and frequency
indices: $ {Q_0}^{\alpha\beta}_{ij} (q;\omega,\omega^\prime) =
Q_0^\alpha (\omega)\, \delta_{q,0}\, \delta_{\omega\omega^\prime}\,
\delta_{\alpha\beta}\, \delta_{ij}, $ where we allow for
$Q_0^{\uparrow}(\omega)\ne Q_0^{\downarrow}(\omega)$. Let us also
assume that the electrons interact through a screened (short-ranged)
interaction. The new saddle-point equation becomes
\begin{eqnarray}
\label{eq:saddle}
Q_0^\alpha(\omega) & = & \frac{i}{\pi \nu_F} \int
\frac{d^2k}{(2\pi)^2}\, G_0^\alpha(\omega,k) -
\frac{2\tau}{\pi \nu_F} \Gamma^{(1)}_1{\Big |}_{Q_0} \nonumber\\ 
& &
-\frac{4\tau}{\pi\nu_F}\, \sum_{\alpha,\beta,\mu} \gamma^{\alpha\beta;\mu\mu} 
\int
\frac{d\omega^\prime}{2\pi}\, Q_0^\mu(\omega^\prime) 
\end{eqnarray}
The first term on the right-hand side of Eq.~(\ref{eq:saddle}) is the
only one present in the non-interacting case. The second term is the
one-loop contribution while the third one corresponds to the
Hartree-Fock approximation. In the non-interacting case the solution
to the saddle-point equations is simply $Q_0^\alpha(\omega) = {\rm
sgn}(\omega)$. For the interacting case, we separate the singlet and
triplet contributions,
\begin{equation}
\gamma^{\alpha\beta;\mu\nu} = \frac{(\pi\nu_F)^2}{2} \left(
\Gamma_s\, \delta_{\alpha\beta}\, \delta_{\mu\nu} + \Gamma_t
\sum_{r=1}^3 \sigma_r^{\alpha\beta} \sigma_r^{\mu\nu} \right),
\end{equation}
where $\Gamma_s$ and $\Gamma_t$ are the singlet and triplet coupling
constants, respectively, and ${\sigma}_r$ are Pauli matrices. While
the singlet channel leads to a shift in the energy band bottom, a
repulsive interaction in the triplet channel causes a magnetic
instability related to a net spin polarization, similar (but not
identical) to the Stoner instability of clean, itinerant electrons in
the presence of a ferromagnetic interaction. In order to capture this
effect, we make the Ansatz
\begin{equation}
Q^\alpha_0(\omega)= \zeta\, {\rm sgn}(\omega) - 2i\tau \Delta_\alpha,
\label{eq:Ansatz}
\end{equation}
where $\zeta$ and $\Delta_\alpha$ have to be determined
self-consistently. For simplicity, we set $\Gamma_s = - \Gamma_t =
\Gamma/2$.

The contribution from the triplet channel to the ${\cal D}= \langle
\delta Q \,\delta Q \rangle$ propagator can be written as
\begin{equation}
\label{eq:dressedD}
{\cal D}(\omega,q) = -\frac{\Gamma (\pi\nu_F)^2}{\zeta^2}\,
D_0^{\alpha\beta}(\omega,q) \, D_2^{\alpha\beta}(\omega,q)
\end{equation}
with
\begin{equation}
D_{0,2}^{\alpha\beta}(\omega,q) = \frac{\zeta^2(2/\pi\nu_F)}
{z_{0,2}|\omega| + Dq^2 + i\, {\rm sgn}(\omega) \Delta_{\alpha\beta}}.
\end{equation}
Here $D$ is the diffusion constant, $\Delta_{\alpha\beta} = \Delta_\alpha -
\Delta_\beta$, $z_0 = z$ (the frequency renormalization factor), and $z_2 =
z + \nu_F\Gamma$. It becomes apparent that $\zeta$ renormalizes
amplitudes, similarly to the wave-function renormalization which
appears in the RG treatment of the problem
\cite{CCLM,Belitz-Kirkpatrick,keldyshpaper}. By dimensional analysis,
we expect the bare interaction coupling constant $\Gamma$ to be
replaced by $\Gamma/\zeta^2$ -- we have already taken that into
account when writing Eq.~(\ref{eq:dressedD}). Thus, the contribution
to $\Gamma^{(1)}_1$ with opposite spins ($\alpha\ne\beta$) in the loop
is
\breakon
\begin{eqnarray}
 \Gamma^{(1)}_1{\Big |}_{Q_0} & \approx & \frac{i}{8\tau^3} \int
 \frac{d^2k}{(2\pi)^2} \int \frac{d^2q}{(2\pi)^2} \int
 \frac{d\omega^\prime}{2\pi} \frac{1}{[i\omega - \epsilon_k + i
 Q^\alpha_0(\omega)/2\tau]^2}\, \frac{1}{i(\omega + \omega^\prime) -
 \epsilon_{k+q} + i Q^\beta_0(\omega + \omega^\prime)/2\tau} \nonumber
 \\ & & \times\, \frac{-4\zeta^2 \Gamma}{\left[ \bar{z}
 |\omega^\prime| + Dq^2 + i\, {\rm sgn}(\omega^\prime)
 \Delta_{\alpha\beta} \right]^2} \nonumber \\ & \approx & -
 \frac{\zeta^2\nu_F^2\Gamma} {8\pi g\bar{z}\tau} \frac{{\rm
 sgn}(\omega)} {(\zeta + i\Delta_{\alpha\beta}\tau)^2} \left[ - \ln
 \sqrt{(\omega\tau)^2 + (\Delta_{\alpha\beta}\tau/\bar{z})^2} + i
 \arctan (\Delta_{\alpha\beta}/\bar{z}\omega) \right],
\label{eq:oneloopterm}
\end{eqnarray}
\breakoff
\noindent 
where $g = \nu_F D$ is the dimensionless conductance in 2-D and
$\bar{z} = (z + z_2)/2$. In Eq.~(\ref{eq:oneloopterm}) we have
constrained $|\omega|, |\Delta_{\alpha\beta}| \ll 1/\tau$. 

With the one-loop contribution Eq.~(\ref{eq:oneloopterm}), one can now
replace it in Eq.~(\ref{eq:saddle}) and solve self-consistently for the
magnetization bandwidth $\Delta = \Delta_{\uparrow}-\Delta_{\downarrow} $ .
After some algebraic manipulations, we find that the $\Delta$ in the
saddle-point solution must satisfy
\begin{equation}
\label{eq:Delta} 
\Delta \approx \frac{\bar{z}}{\tau} \exp \left[ - \frac{(2\pi\bar{z})^2
g} {2\nu_F\Gamma} \left( \frac{1}{\nu_F\Gamma} - 1 \right) \right].
\end{equation}
The upper bound in frequency of the diffusion propagator
sets the prefactor of the exponential in Eq.~(\ref{eq:Delta}). As a
result, $\Delta$ is proportional to the elastic scattering rate,
$1/\tau$, rather than the total bandwidth or the Fermi energy.

Equation~(\ref{eq:Delta}) points to the existence of a non-zero spin
polarization $\Delta>0$ for {\it any} positive value of $\Gamma$, provided
that the dimensionless conductance is finite. Only for infinite $g$ we
recover the usual Stoner instability characteristic of clean systems, namely,
a ferromagnetic instability at $\Gamma > 1/\nu_F$. In the particular case of
a {\it finite} system, the same tendency towards spin polarization was found
by Andreev and Kamenev \cite{AntonAlexPRL}. This suggests that the
ferromagnetic instability may be a robust property of 2D disordered
interacting electrons, since their starting point was rather different than
ours. For the finite system case, they used an exact basis representation for
the non-interacting problem, combined with disorder averaged Hartree-Fock
matrix elements. These matrix elements were dressed by the diffusive dynamics
to lowest order in $1/g$ and provided a contribution to the magnetization
similar, but not identical, to our one-loop calculation with the disorder
average carried out from the very beginning.

It has been suggested that the divergence of the triplet coupling is
connected to the formation of local magnetic moments in the system
\cite{Milo-Sachdev-Bhatt,Bhatt-Fisher}. Notice, however, that the
polarization of the saddle-point solution in Eq.~(\ref{eq:Delta}) connects
continuously with the Stoner instability in the limit of a clean system. This
is suggestive of a dirty ferromagnetic state\cite{fradkin} where there exists
a residual coupling between the local moments, causing a tendency towards a
global ferromagnetic order at $T=0$. If the localized moments are coupled
through, for example, an RKKY interaction that has alternating signs, one
would expect not a ferromagnetic state, but rather something similar to a
spin glass state \cite{Aleiner}. There can be two solutions to this puzzle.
One is that the characteristic size of the local moment diverges, and hence
the system is in a ferromagnetic state (at $T=0$) according to the
saddle-point solution in this paper. The other alternative is that there may
be another saddle, which in contrast to ours, breaks replica symmetry (RS).
Let us remind that in spin glass models the replica symmetric solution is
unstable in the glass phase, which is characterized by a non-zero
Edwards-Anderson order parameter without magnetic order. Our saddle-point
solution, on the other hand, finds directly a magnetic order parameter
$\Delta\ne 0$. In this case, it appears unlikely that RS breaking may resolve
the issue. In addition, any RS breaking solution must be such that it
recovers the usual Stoner instability in the clean limit.

Another important issue to consider is the divergence of the frequency
rescaling factor $z$ in the RG equations for the sigma model expanded
around the zero-loop saddle. In the introduction, we claimed that the
divergence of the interactions happen at a finite length scale;
however, the divergence of $z$ would mean that the associated energy
scale goes to zero. The latter conclusion is not correct for the
following reason. The coupling constant with the leading divergence is
the triplet, and when this coupling leaves the perturbative RG limit,
one can no longer trust the flow. The saddle with the non-trivial
magnetization resolves the leading triplet divergence, and that should
stop the $z$ divergence as well.

We would like to conclude the paper by discussing the issue of localization
in disordered and interacting systems. The nature of the triplet channel
instability is clearly captured by a disordered averaged effective theory
through a non-trivial saddle-point solution with a non-zero magnetization.
Moreover, the saddle point provides a natural starting point of an new RG
treatment for the localization problem.  Once $\Delta\ne 0$ is taken, it
appears that the renormalization of the diffusion constant $D$ should proceed
similarly to the non-interacting case.  Although a renormalization program
should be carried out to confirm whether $D$ scales to zero, it appears
unlikely that starting from the new saddle will lead to delocalization in 2-D
for weak interactions.  This, however, would not preclude a correlated
(super) conducting state, for example, as proposed by Phillips and coworkers
\cite{everybody}.


We are grateful to E. Fradkin and P. A. Lee for valuable
comments. Support was provided by the NSF Grant DMR-98-76208, the
Alfred P. Sloan Foundation (C.~C.~C), and the Brazilian agencies CNPq,
FAPERJ, and PRONEX (E.~R.~M.). The authors also thank the Aspen Center
for Physics for the hospitality at the initial stages of this work.


\end{multicols} 

\end{document}